\documentclass[twocolumn,showpacs,preprintnumbers,amsmath,amssymb]{revtex4-1}

\usepackage{graphicx}
\usepackage{dcolumn}
\usepackage{bm}
\usepackage{amsfonts}
\usepackage{amsmath}
\usepackage{subfigure}
\usepackage{epsfig}
\graphicspath{{./}{./FIGS/}{./data/}}

\begin{document}


\title{Cutting-Plane Algorithms and Solution Whitening for the Vertex-Cover Problem }
\author{G. Claussen$^1$}
\email{gunnar.claussen1@uni-oldenburg.de}
\author{A. K. Hartmann$^1$}
\email{alexander.hartmann@uni-oldenburg.de}
\affiliation{
$^1$ Institut f\"ur Physik, Universit\"at Oldenburg, Carl-von-Ossietzky-Stra{\ss}e 9--11, 26111 Oldenburg, Germany\\
}

\date{\today}


\begin{abstract}
The phase-transition behavior of the NP-hard vertex-cover (VC) combinatorial optimization problem is studied numerically  by linear programming (LP) on ensembles of random graphs. As the basic Simplex (SX) algorithm suitable for such LPs may produce incomplete solutions for sufficiently complex graphs,
the application of cutting-plane (CP) methods is sought. We consider \emph{Gomory} and $\{0,\frac{1}{2}\}$ cuts. We measure the probability of obtaining complete solutions with these approaches as a function of the average node degree $c$ and observe transition between typically complete and incomplete phase regions. While not generally complete solutions are obtained for graphs of arbitrarily high complexity, the CP approaches still advance the boundary in comparison to the pure SX algorithm, beyond the known replica-symmetry breaking (RSB) transition at $c=e\approx 2.718$. In fact, our results provide evidence
for  another algorithmic transition at $c\approx 2.90(2)$.
 Besides this, we
  quantify the transition between \emph{easy} and \emph{hard} solvability of the VC problem also in terms of numerical effort. Further we study the so-called whitening of the solution, which is a measure for the degree of freedom that single vertices experience with respect to degenerate solutions. Inspection of the quantities related to clusters of white vertices reveals that whitening is affected, only slightly but measurably,  by the RSB transition.
\end{abstract}

\maketitle


\section{Introduction \label{sec:introduction}}

Phase transitions in random ensembles of combinatorial optimization problems \cite{Hartmann2005,mezard2009,moore2011} have  been attracting the statistical physics community since more than two decades. In particular so called \emph{easy-hard} transitions have been observed, where for a control parameter of the random ensemble there exist critical values such that, for a given algorithm, on the \emph{easy} side a problem can be typically solved in polynomial time, while in the \emph{hard} region an exponential effort is necessary. Phase transitions on suitably chosen ensembles of random instances were found, e.g., for the Satisfiability Problem (SAT) \cite{kirkpatrick1994}, the Traveling Salesperson Problem 
(TSP) \cite{gent1996} or the vertex-cover problem (VC) \cite{cover2000}. Note that such transitions describe the relationship between optimization problems and algorithms to solve them. In the physics community, so far algorithms have predominately studied which are based on moving  in or representing the space of feasible configurations. These algorithm  are exact branch-and-bound algorithms \cite{davis1962,sedgewick1990}, stochastic algorithms, like WalkSAT \cite{papadimitriou1991} or ASAT \cite{ardelius2006} and  message-passing algorithms \cite{mezard2002,mezard2002b}, which are inspired by statistical mechanics methods like the cavity approach \cite{mezard1987}. Instead, we follow here a different approach, namely \emph{linear programming} (LP) \cite{papadimitriou1998} in connection with \emph{cutting planes} (CP) \cite{cook1998}. Such approaches operate \emph{outside} the space of feasible solutions for most of the computation time. Although for practical applications many CP implementations exist, e.g., based on commercial packages like CPLEX \cite{CPLEX}, the phase transition behavior with respect to easy-hard transitions has not been studied widely. Recently and
 for the first time to our knowledge, a phase transition in a problem-specific CP approach was observed and analyzed \cite{vc_lp2012}. Based on this, we perform here a corresponding study for two very general CP approaches, so-called Gomory and $\{0, \frac 1 2 \}$ cuts, for the VC problem on Erd\H{o}s-Renyi (ER) random graphs.
  
The VC problem describes the intricate task of finding the vertices (also called nodes) of a graph that make up a subset called \emph{cover} in such a way that all edges of the graph are adjacent to at least one vertex in the cover set. It forms a prime example of a combinatorial optimization problem\cite{papadimitriou1998}, as the decision whether one vertex is part of the cover set or not is Boolean. As many problems of applications can be mapped onto graphs, it comes as no big surprise that also the VC problem has been applied to model real-world problems such as the number of guards needed in a museum \cite{cover2000}, the stationing of police cars on road networks \cite{Adler2014}, the placement of sensor devices in wireless communication networks \cite{Safar2007}, vaccination strategies against disease spreading \cite{Boros2007}, or even cooperative robot surveillance \cite{Parker1997}. 

Previously the phase transition behavior of VC was studied mostly \cite{cover2000,cover-time2001,cover-long2001,hclg2003} in connection with algorithms based on the space of feasible solutions. For the case of ER random graphs, it turned out that the connectivity $c_{\text{c}}=e\approx 2.718$, i.e. the Eulerian number plays a crucial role. For $c<e$, VC is typically easy \cite{bauer2001} and
can analytically be solved by a replica-symmetric approach \cite{cover2000}, hence the solutions landscape is dominated by one single cluster in the thermodynamic limit. For $c>e$, replica-symmetry breaking (RSB) appears, as visible by many clusters in the solution landscape \cite{vccluster2004}, and the problem
 appears typically hard to solve by exact algorithms. Nevertheless, VC can be comfortably expressed as an LP, where variables $x_i=0,1$ describe whether node $i$ belongs to the cover subset or not, see below for a technical description. With this, a \emph{relaxed} LP, i.e., with $x_i \in [0,1]$, can be solved through the Simplex (SX) algorithm \cite{thie2008}. A simple measure of the complexity of a graph is
the connectivity $c := 2\frac{M}{N}$, where $M$ is the number of edges and $N$ the number of vertices. As the VC problem is noterministic polynomial (NP)-hard \cite{Karp1972}, with increasing complexity of the graph, the plain SX algorithm fails to produce correct (also called \emph{complete}) results with unambiguously decided variable values $x_i \in \{0,1\}$. Instead the yielded solutions include an increasing amount of undecided or fractional values $x_i \in [0,1]\setminus \{0,1\}$. Such solutions are called \emph{incomplete}, and generally the fraction $p$ of complete solutions obtainable with the SX algorithm breaks down when increasing $c$. This represents, e.g. on the ensemble of random graphs, such 
an aforementioned phase transition between an ``easy'' and a ``hard'' optimization problem \cite{vc_lp2012}.

The SX algorithm can be improved by the introduction of cutting planes, i.e., sets of linear inequalities which reduce the volume of the space of possible solutions. For the study of phase transitions of VC, first problem-specific cutting planes were investigated, based on the topology of the problem \cite{vc_lp2012}. There, it was reasoned that for closed cycles within a graph that consist of an odd number of vertices, the respective number of vertices covered within this cycle has to be equal to the the rounded-up half of the number of vertices. With this, additional constraints could be entered into the LP. With these measures a transition regarding the hardness of the problem was observed at a the connectivity of $c_{\text{c}}=e$ that has already been described earlier.

While this approach was so to say tailored for the VC problem and its prerequisites, it is tempting to utilize more general approaches as well. A well-established one is the so-called Gomory cut, which produces new constraints from a given solution by regarding only the fractional parts of the coefficients in the SX \emph{tableau} \cite{Gomory1958}. Originally is was viewed as a theoretical way only, also because of arising numerical instabilities \cite{parker1988}. Nevertheless, more recently it has since become more widely employed \cite{Cornuejols2007,Gomory2007} for practical optimization, especially in connection with branch-and-cut methods. Note that the aim of computer scientists is an engineering one, to provide as efficient algorithms as possible, typically tested for test-beds of problems, and leading usually to implementations of clever combinations of algorithms. In the present work, we are more interested in the behavior of certain isolated algorithms when applied to ensembles exhibiting easy-hard transitions. This means we are interested in the statistical mechanics properties and the 
relation to the computational complexity of certain algorithms, even if in some cases the algorithmic performance is bad.

A similar CP method is the $\left\{0,\frac{1}{2}\right\}$ (zero-half) cut, which is a special case of the so-called Chv\'atal-Gomory cut \cite{Caprara1996}. Within this approach, linearly independent rows of this tableau are added up until an odd number of coefficients is either equal to 0 or can be divided to $\frac{1}{2}$ respectively, hence the name. The right-hand side of the
inequality can then be rounded up or down (depending on the sign) to the next integer. Note that these cuts are also noted to combine ``cyclic'' inequality constraints \cite{Yuceoglu2015}, thus the cycle-cut scheme employed by \cite{vc_lp2012} can just be interpreted as an applied case of zero-half cuts.

For the present work, we again start by solving the VC of ER graphs by application of the plain SX algorithm, but we perform a thorough analysis of the results, which was lacking in previous work. Next, we expand the SX algorithm by allowing Gomory cuts, zero-half cuts or both combined. For comparison, we also employ an exact algorithm, but restricted in system size.
We investigate the criticality of the VC problem's hardness with respect to the fraction of complete solutions, the fraction of covered vertices and the deterministic calculation time needed for obtaining the solution.

A further property we put under scrutiny in the present work is the so-called \emph{whiteness} of the solution. This is not directly related to cutting planes approaches, but gives also some insight into the solution landscape and has been, to our knowledge, not been studied before for VC. As the name indicates, the method was originally proposed for graph-coloring problems \cite{Parisi2003} and more prominently applied to the $K$-SAT problem later \cite{Braunstein2004,Seitz2005}. Whitening was used to identify, e.g., clusters of solutions that differ in only one variable \cite{Braunstein2004,Maneva2007}. A detailed study on the more general $K$-SAT problems was presented later on \cite{Alava2008,Braunstein2016}. Whitening procedures have been used also in the calculation of SAT backbones \cite{Zhang2018}. The interesting property to put under particular scrutiny with respect to solutions of the VC problem is the dependence of the ``whiteness'' of the solution on the solvability of the respective problem. In the literature on whitening, this idea has been presented in variousways. First, there is the notion of a ``freezing'' transition: With an increased number of clauses, i.e. constraints in the SAT problem, the fraction of ``frozen'' solutions rises, i.e., the fraction of those variables which exhibit in all degenerate solutions the same value. For a low density of constraints, the solutions are nearly always ``white'', also called ``unfrozen'' or ``free'' in other references \cite{Ardelius2008,Braunstein2016}. Thus, it has been assumed that the presence of white solutions indicates easy solvability \cite{Alava2008}. A particular mark of this property is the fact that the average whiteness depth (AWD), which expresses the average iteration number of the whitening procedure in which a variable of the model is marked as white, remains finite. It appears however that this property has not been inspected any further afterwards.

Note that the whitening procedure is not applicable to VC solutions in its original form, but we present below a corresponding adjusted version. In the results section, we will investigate quantities related to the prevalence of white vertices in the graph and interconnected clusters of these.



This paper is outlined as follows: Section \ref{sec:model} introduces the VC problem formally and its LP formulation. Next, in Secs.~\ref{sec:cutting-plane_algorithms},\ref{sec:whitening-alg} and \ref{sec:analysis} we explain the numerical algorithms we have used. In section \ref{sec:results} we show our results including data for the typical computaional hardness. Ultimately, Sec.~\ref{sec:conclusions} sums up the results briefly and gives an outlook to further desirable research on this topic.


\section{Methods}

We first introduce the formal definition of the vertex-cover problem and show how it can be written as a linear program. Next, we explain the CP approach. In the third section, we show our whitening procedure as adapted to VC. Finally, we define the finite-size scaling approach we have used to analyze the data.

\subsection{The vertex-cover problem as a linear program \label{sec:model}}

Generally, the vertex-cover problem is defined on a \emph{graph} $G = (V,E)$, where $V$ is a finte set of $N$ \emph{vertices} (or \emph{nodes})  which are connected by $M$ undirected \emph{edges} $\{i,j\}\in E \subset V^{(2)}$. In turn, the edge $\{i,j\}$ is called \emph{incident} to the nodes $i$ and $j$, and these two edges are called \emph{adjacent} to each other.

Now, the VC problem on $G$ consists of the following: Find a subset $V_C \subseteq V$ such that $\forall \{i,j\} \in E: i \in V_C \vee j \in V_C$. To put it less formal: Each edge present in $E$ should be incident to at least one vertex in $V_C$. The vertices present in $V_C$ are called \emph{covered}, and the subset $V_C$ comprises the \emph{vertex cover} of the graph $G$. Note that with the aforementioned condition, it is possible for an edge to be incident to two vertices of $V_C$.

Trivially, $V_C = V$ is already a vertex cover, but obviously a very large one. The main point in the VC problem is therefore to minimize the size $|V_C|$ of the vertex cover, which is an optimization problem. It belongs to the class of NP-hard problems \cite{garey1979}, which means that currently only exact algorithms are known which require a worst-case running time which grows exponentially in the number $N$ of nodes. This directly motivates the expression of the VC problem as a linear program (LP). In this, all vertices are represented as variables $x_i$ where $x_i=1$ means node $i$ is covered
while $x_i$ means $i$ is not covered. Now, we have the program statement

\begin{align}
 \text{minimize } \sum_{i \in V} x_i \label{eq:VC_LP_objective}   \\
 \text{subject to: } x_i + x_j \geq 1 \: \forall \{i,j\}
 \in E \label{eq:VC_LP_constraints} \\
 x_i \in \{0,1\}\,. \nonumber
\end{align}

Such a scheme is called \emph{integer linear program} (ILP). Eq.~\eqref{eq:VC_LP_objective} denotes the \emph{objective function} of the program. The total sum of variables is to be minimized, i.e. the number of elements $x_i$ in the vertex cover $V_C$. But this minimization has to fulfill the expressions stated in Eq.~\eqref{eq:VC_LP_constraints}, which are called \emph{constraints}: Each edge $\{i,j\}$ present in $E$ is represented through an inequality, and this inequality is fulfilled if the variables corresponding to the vertices $i$ and $j$  sum up to at least 1, i.e., if at least one node is covered.

But unfortunately, such ideal results are only possible if the solution space is integer, i.e. $\vec{x} \in \{0,1\}^N$. With real values
$x_i \in \mathbb{R}$, which makes the problem feasible for polynomially-running LP solvers. One calls the constraint $x_i \in \{0,1\}$
\emph{relaxed}, e.g., to $x_i \in [0,1]$. However, the problem now arises that a constraint may also be fulfilled by e.g. $x_i = x_j = 0.5$.
For example for a triangle graph, i.e., the complete graph with three nodes, $x_1=x_2=x_3=0.5$  is the optimum solution of the relaxed problem,
while a true optimum VC solution has $\sum_{1}^3 x_i=2$. Obtaining such a non-integer solution, in particular for large
problems, gives no clear deciding statement whether $i$ and $j$ are actually part of the vertex cover. Solutions of the LP with such an outcome on any of the variable values $x_i$ are called \emph{incomplete}.

Note that actually solving ILPs is a research topic of its own, and has spurred the development of many now-common methods. Most prominent among these are tree-based \emph{branching} methods \cite{Dakin1965} and CP approaches \cite{Kelley1960}. Such approaches have an exponential worst-case running time. The latter approach and its phase transition behavior with respect to typically polynomial running times is the main focus of this work.


\subsection{Cutting-plane algorithms \label{sec:cutting-plane_algorithms}}

To explain the approaches we used, we have to resort a bit to the theory of linear programming, for easy introductions see \cite{papadimitriou1998,thie2008}. Nevertheless, we will mention only those elements of LP, which are needed here.

We start considering the relaxed VC problem, i.e., with $x_i \in [0,1]$, the space $\vec{x} \in [0,1]^N$ is the $N$-dimensional unit hypercube. The
linear inequalities in Eq.~\eqref{eq:VC_LP_constraints} constitute dividing lines in the $x_i$-$x_j$ plane. Thus, the relaxed solution space becomes limited in nearly all directions, and instead of a hypercube, it is now equivalent to a polytope.

The linearity of the optimization problems means that for any points inside the polytope there is a direction in which one can improve the objective function. Therefore, optimum solutions are found on the vertices or facets of the polytope. To apply the SX algorithm, the problem is slightly
rewritten, by replacing the inequalities $x_i + x_j \geq 1$ by equalities $x_i+x_j-s_m=1$ with an additional \emph{slack variable} $s_m \le 0$  for each constraint. Note that for VC, the constraints imply $s_m\in[0,1]$. Since there are $M$ edges, there are $M$ constraints and therefore $m=1,\ldots,M$ slack variables $s_m$. This set of equations is written in matrix form as 

\begin{align}
 A  \vec{x} = \vec{b}\,, \label{eq:LP_constraints_general}
\end{align}

with $A$ being an $M \times (N+M)$ matrix describing the equalities. For simplicity, we denote the combined vector of the problem variables $x_1,\ldots,x_N$ and the slack variables $s_1,\ldots, s_{M}$ as the $(N+M)$ dimensional vector $\vec{x}$. Note that here $\vec{b}=(1,1,\ldots,1)^T \in \mathbb{R}^M$ and that $A$ has rank $M$ since all edges are independent in the graph.

As the theory of LP \cite{papadimitriou1998,thie2008} tells us, SX works by selecting so called \emph{basic  feasible  solutions}. These are subsets ${\cal B}=\{{\cal B}(1),\ldots,{\cal B}(M)\}$ of indices ${\cal B}(i) \in \{1,\ldots, N+M\}$ such that the columns of $A$ corresponding to these indices are linearly independent. Thus, these columns form a $M\times M$ matrix of rank $M$ denoted as $B$. The other indices of the other columns of $A$ are denoted as ${\cal Z}$, forming the matrix $Z$. W.l.o.g., let the columns and rows of $A$ and the corresponding 
entries of $\vec{x}$ and $\vec{b}$ be ordered in such a way that $A=(B,Z)$ and $\vec{x}=(\vec{x}_B,\vec{x}_Z)^T$. Thus, Eq.~(\ref{eq:LP_constraints_general}) can be written as $A  \vec{x} = B\vec{x}_B+Z\vec{x}_n= \vec{b}$. Since $B$ has rank $M$, the inverse matrix $B^{-1}$ exists, and one obtains $ B^{-1}B\vec{x}_B+B^{-1}Z\vec{x}_Z= B^{-1}\vec{b}$, i.e., $\vec{x}_B+B^{-1}Z\vec{x}_Z= B^{-1}\vec{b}$. By selecting $\vec{x}_Z\equiv 0$ one obtains the basic feasible solution 

\begin{equation}
\vec{x}_B= B^{-1}\vec{b}\,,\quad \vec{x}_Z\equiv 0\,.
\label{eq:bfs}
\end{equation} 

We just mentioned that SX works by first determining an initial basic feasible solution, which is sometimes involved, and then exchanging variables in and out of ${\cal B}$ while assuring that the value of the objective function is not changed opposite to the desired direction. This is interated until no further improvement is possible, i.e., the optimum is found, which also fulfills Eq.~(\ref{eq:bfs}).

Now, since the problem is relaxed, i.e., $\vec{x}\in [0,1]^{N+M}$, there may be non-integer entries of $\vec{x}$ such that the original ILP Eqs.(\ref{eq:VC_LP_objective}),(\ref{eq:VC_LP_constraints}) is not solved. Note that if the relaxed problem leads actually to an integer, i.e., complete solution, it is automatically a solution of the ILP and we are done.

Hence, we now consider the case of an incomplete solution, i.e., there are entries in $\vec{x}_B$ which are, for our problem, in $]0,1[$. We next explain so called \emph{Gomory cuts} \cite{Gomory1958,Gomory2007}, which have the basic idea to generate inequalities which make the present incomplete version invalid, while not affecting any solution to the ILP. The starting point is to multiply Eq. (\ref{eq:LP_constraints_general}) with $B^{-1}$ leading to

\begin{equation}
 B^{-1} A  \vec{x} = B^{-1} \vec{b}\,. 
\end{equation}

Following in our presentation\cite{gade2013}, we denote the entries of $B^{-1}A=$ by ${\bar a}_{ij}$, and the entries of $B^{-1} \vec{b}$ by $\bar b_i$. Note that after again suitable reordering or rows and columns, $B^{-1}A=B^{-1}(B,Z)= (E,B^{-1}Z)$, where $E$ is the unit matrix. According to Eq. (\ref{eq:bfs}) the right hand side just contains the non-zero entries  of the solution of the relaxed LP, i.e., of the variables belonging to the optimum basic feasible solution. Since this solution is not integer, there must be at least one line $i$ where the right hand side is non-integer. This line reads:
\begin{equation}
x_{{\cal B}(i)} + \sum_{j \in {\cal Z}} \bar{a}_{ij}x_j = \bar b_i\,.
\label{eq:gomory:A}
\end{equation}
Since all variables fulfill $x_i\ge 0$, by rounding up the coefficients on the left hand side to $\lceil \bar a_{ij} \rceil$ we immediately get $x_{{\cal B}(i)} + \sum_{j \in {\cal Z}} \lceil \bar{a}_{ij} \rceil x_j \ge \bar b_i$. If all variables $x_i$ on the left hand side, also the added slack variables, are required to be integer in the end, the full left-hand side must be integer. Thus, we can round up the right hand side as well, resulting in
\begin{equation}
x_{{\cal B}(i)} + \sum_{j \in {\cal Z}} \lceil \bar{a}_{ij} \rceil x_j \ge \lceil \bar b_i \rceil\,,
\label{eq:gomory:B}
\end{equation}

For the current solution $x_i=0$ for $j\in {\cal Z}$ and $\bar b_i$ is non-integer, which contradicts Eq.~\eqref{eq:bfs}. Thus, the inequality is not fulfilled by the current solution. Such an inequality is called \emph{Gomory cut}.

If not all coefficients $\bar a_{ij} \in \mathbb{Z}$, then one can equivalently subtract Eq.~(\ref{eq:gomory:A}) from Eq.~(\ref{eq:gomory:B}). By denoting the missing part of a real number $r$ to its next integer as $\phi(r)=\lceil r \rceil -r$, we arrive at

\begin{equation}
\sum_{j \in {\cal Z}} \phi(\bar{a}_{ij}) x_j \ge \phi(\bar b_i )\,.
\label{eq:gomory:C}
\end{equation}

For the current solution $x_i=0$ for $j\in {\cal Z}$ and $\bar b_i$ is non-integer, thus $\phi(\bar b_i )>0$. Thus, also Eq.~(\ref{eq:gomory:C}), which is called \emph{Gomory fractional cut}, is \emph{not} fulfilled by the current solution. 

With the addition of the the Gomory cut given by Eq.~\eqref{eq:gomory:A} or Eq.~\eqref{eq:gomory:C}, the LP can be re-solved. Afterwards, theroblem either yields a complete solution or allows additional feasible Gomory cuts, and this scheme is iterated until either a complete solution is obtained or no more Gomory cuts can be generated. 

The main notion behind the usage of Gomory cuts is, in the words of \cite{papadimitriou1998}, that ``no integer feasible points are excluded'', while the solution space is still curtailed in a meaningful way, therefore generally leading towards the correct, optimum solution of the LP. It can be proven \cite{gade2013} that in principle the iteration of adding Gomory cuts will terminate after a finite number of steps in an integer solution, if it exists. Here, in particular the so called \emph{lexicographical dual} SX algorithm yields a good performance \cite{Zanette2011}. As mentioned before, for practical implementation it is at least possible that no further Gomory cuts be generated at some point -- which then might terminate the execution of the algorithm without finding a complete solution. It has also been argued that Gomory cuts are generally prone to numerical instability \cite{Cook2009}, due to computational problems in floating-point arithmetic when representing fractional numbers and possibly also due to the rapid expansion of the LP size generated by the addition of constraints and slack variables. A possible remedy of the first problem could be to treat all constraint coefficients $a_{ij}$ generally as fractions instead of floating-point numbers.

The other considered CP method are $\{0,\frac{1}{2}\}$ (zero-half) cuts. In order to understand this approach, we note that the Gomory cut Eq.~\eqref{eq:gomory:B} has the form 

\begin{equation}
  \lceil \vec\lambda^T  A \rceil x \geq \lceil \vec \lambda^T
 \vec b\rceil \,,
\label{eq:Chvatal-Gomory_cut}
\end{equation}

where the vector $\vec\lambda \in \mathbb{R}^m$ originates from the corresponding entries of the inverse matrix $B$. Still, this holds for arbitrary vectors $\vec\lambda$, starting from $A\vec{x}=\vec b$ and also from $A\vec x\ge \vec b$, if one uses the original starting LP Eq.~\eqref{eq:VC_LP_constraints} without slack variables, with the same argument when deriving Eq.~\eqref{eq:gomory:B} above. If Eq.~\eqref{eq:Chvatal-Gomory_cut} excludes the current optimal but non-integer solution, it can be useful for a CP scheme. For the $\{0,\frac{1}{2}\}$ cuts, one restricts oneself to vectors $\vec{\lambda}\in \{0,\frac 1 2\}^m$. Note that useful inequalities arise only if actual rounding takes place, this means that in $A$ and $\vec{b}$ odd entries must exist. But this can be guaranteed always, because if all entries in a line are even, one can divide by two as many times as needed until at least one entry becomes odd, without changing the meaning. Although the set of possible vectors $\vec{v}$ is exponentially large, there exist efficient methods to find cuts which are violated by the current non-integer solution \cite{Caprara1996,koster2009}. Since we use an external library for our simulations, we do not go into details here.

In the introduction, we mentioned that the earlier CP approach to the VC problem of \cite{vc_lp2012} is a special case of the $\{0,\frac{1}{2}\}$ cut. This is to be understood in the following way: In the VC problem, the basic constraints as per Eq.~\eqref{eq:VC_LP_constraints} include only two variables, i.e., those denoting whether the respective vertices incident to the respective edge are covered or not. If one now considers a cycle ${\cal L} \subset E$ in the graph and sums up the inequalities corresponding to the edges in the cycle, each node of the cycle will contribute its corresponding variable twice, leading to $2 \sum_{i:  \exists e \in {\cal L}: i \in e} x_i \ge |{\cal L}|$. After multiplying with $\frac 1 2$ and rounding up, which may happen only at the right side, one arrives at nontrivial inequalities for odd cycle length $|{\cal L}|$. Therefore, the application of the $\{0,\frac{1}{2}\}$ CP approach is a generalization of the cuts used before \cite{vc_lp2012} allows for a comparison with the previous results.

Regarding the actual implementation of those CP methods we used the IBM ILOG CPLEX optimization studio \cite{CPLEX}. In CPLEX, we use the pre-defined methods supplied by CPLEX when defining the LP specifically as an ILP. This is done by changing the type of all present variables from floating-point numbers to integers. For some our our results shown below, we have used the full power of CPLEX for the solution of the ILP. But for our analyses of the effects of the cutting planes, we have deactivates manually other other cut types except Gomory cuts and zero-half cuts. We then deactivated also branch-and-cut methods, which is done by limiting the number of branching nodes to zero, and all other the heuristic methods, which are anyway not further specified by the CPLEX documentation.

For some performance tests and comparison, in particular for counting the number of generated cutting inequalities, we also used our own implementation of Gomory cuts while using CPLEX just as LP solver. Nevertheless, we observed the issues mentioned earlier regarding numerical stability, i.e., the difficulty of representing fractional values like $\frac{1}{3} = 0.\overline{3} = 0.333...$ with sufficient precision. Thus, we have set a maximum number $N_{\rm cut,\max}=1000$ of added CPs. If  this threshold was reached, without finding a complete solution, the corresponding instance was counted as incomplete.

\subsection{Whitening\label{sec:whitening-alg}}

We start by considering the whitening approach for $K$-SAT \cite{Braunstein2004,Seitz2005}. A SAT instance is a conjunction of logical clauses, where each clause contains disjunctions of possibly negated variables called \emph{literals}. For $K$-SAT each clause contains $K$ literals. The SAT problems is a decision problem which asks whether for a given instance there exists a satisfying ``true'' assignment. Whitening is meant to distinguish white and frozen variables within a solution, where flipping white variables affects the solution only locally, but not globally. Therefore, the whitening algorithm for SAT start by assigning the white state to all clauses which are satisfied by at least two literals, i.e., at least one variable can be flipped without changing the satisfied state of the clause. Next, iteratively, variables are assigned white which appear as satisfying only in white clauses. At the same time, more clauses can turn white if they contain a white variable. The iteration stops if no changes occur.

Unfortunately, this algorithm cannot be directly transferred to VC: The clauses of SAT correspond to the edges of VC. Therefore, in the initial phase, a white state of an edge would be assigned if the edge is covered twice. Now, a node would be considered white if all its incident edges are white. But this would mean that the node and all its neighbors are covered, which is a contradiction to the minimum property of a minimum vertex cover.

Thus, we have adapted the approach for the VC problem. The basic idea is that for a given minimum vertex cover, a covered vertex $i$ which has at most one uncovered neighboring vertex $j$, one can cover $i$ instead of $j$. Thus the vertex $i$ can be in two states, covered and uncovered, which is considered as white. Corresponding cases hold for white neighbors. The actual algorithm reads as follows:

\begin{itemize}
 \item Given: a vertex cover $V_C$
 \item Initialization: Mark every vertex as frozen, i.e., non-white.
 \item Repeat until no vertex state changes:
 \begin{enumerate}
  \item Mark all vertices $i$ white that are covered ($i \in V_C$) and 
that have a maximum of \textbf{one} neighbor which is non-covered and
 frozen.
  \item Mark all vertices $i$ white that are not covered ($i \notin V_C$), but adjacent to white vertices only
 \end{enumerate}
\end{itemize}

\begin{figure}[ht]
 \includegraphics[width=0.2\textwidth]{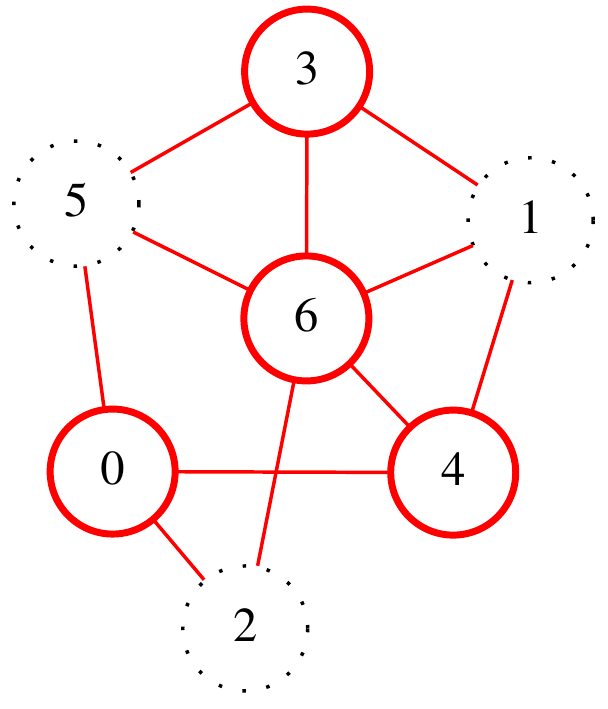}
 \includegraphics[width=0.2\textwidth]{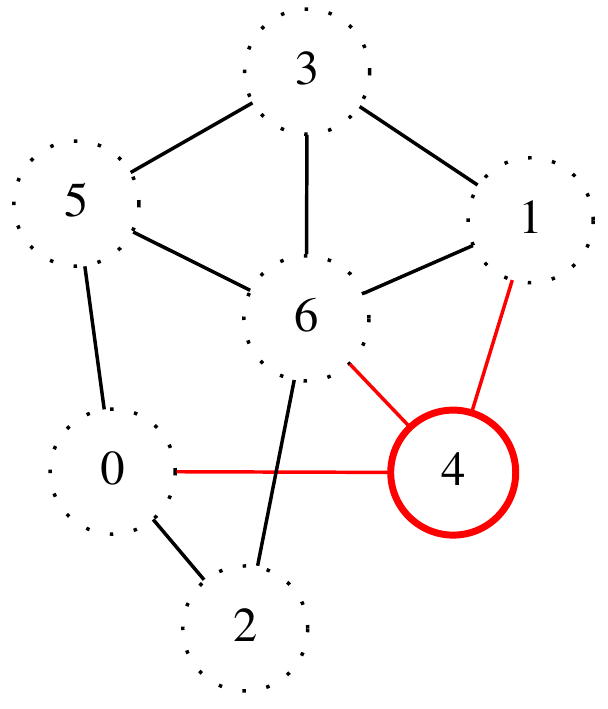}
 \caption{(color online) Example graph with minimum vertex cover (left). Vertices with solid lines indicate covered vertices dotted lines indicate uncovered ones. (Right) Result of the whitening algorithm: Solid circles represent white nodes, frozen ones are shown with dotted lines.}
 \label{fig:example_whitening}
\end{figure}

The notion behind this scheme is exemplified by the application shown in Fig.~\ref{fig:example_whitening}: In the first step, a single node could be declared as white, i.e., node 4. This is due to the degree of freedom attributed to this vertex in the solution. Flipping vertex 4 would leave only edge $(1,4)$ uncovered, while the other edges incident to vertex 4, i.e. $(0,4)$ and $(4,6)$ are also incident to other covered vertices. Hence, a complete solution after flipping vertex 4 can still be obtained by covering vertex 1. The solution $\tilde{V}_C = \{0,1,3,6\}$ is then of the same size $N_C = 4$ as the original solution $\tilde{V}_C = \{0,3,4,6\}$ and hence a degenerate case. For the example graph and the given solution, not more white nodes are detected.

Note that the whitening algorithm is able to gather information about the solution space structure, but it will typically not have access to all degenerate solutions. For the example shown in Fig.~\ref{fig:example_whitening}, there are more solutions: As mentioned, one could cover node 1 instead of node 4. If node 1 is covered, one can cover node 5 instead of node 3. This solution is not detected by the whitening algorithm. We have investigated other variants of algorithms for the whitening, with more relaxed options for a node to become white, e.g., make a node white if all neighbors are either white or covered. In the present example in  Fig.~\ref{fig:example_whitening}, node 1 would immediately turn white. This holds also for more nodes. After some iterations even node 6 would become white as well, which clearly makes no sense since it is covered in all minimum vertex covers. All the variants we have tried, although they looked promising on the first sight, lead to such undesirable results. 

A particular question regarding the whitening procedure is the treatment of isolated vertices of the graph, i.e., those without neighbors. One might consider them frozen, as they do not play any role for the VC solution itself since they are never covered at all. However, they fulfill condition (2) of the whitening procedure if the expression ``adjacent to white vertices only'' is interpreted as ``of all their neighbors (zero), all (zero) are white''.  With defining isolated vertices as white it is also possible to achieve full whitening, i.e., all vertices are white. Thus, we find it justified to define isolated vertices as white here.

\subsection{Scaling analysis \label{sec:analysis}}

Our main interest is to determine phase transitions between regions where our algorithms result in complete solutions, i.e., where all variables are integer, and regions where no or few complete solutions are found. For this purpose we measure the fraction $p$ of graphs with a complete solution as a function of the control parameter $c$, which is the average vertex degree here.

In order to determine the phase transition from the data we follow the scheme established by \cite{Schawe2019} for the 2-SAT problem. Note that this approach works in particular for the case when the transition is not visible by an intersection of the curves for different sizes $N$, but instead on one side of the transition, the data is almost size independent, as it was observed previously for other phase transitions in combinatorial optimization problems \cite{leone2001}. The scaling approach works in the following way. For each number of vertices $N$ of the graph, we follow the course of $p(c)$, seek out the point of the steepest decrease in the curve, align a tangent to it and calculate the intercept $c_0(N)$ with the $c$-axis. The behavior of $c_0(N)$ follows \cite{Schawe2019} the finite-size behavior of other standard phase transitions well, i.e., a power-law behavior as 

\begin{equation}
 c_0(N) = a\cdot N^{-b} + c_{0,\infty} \label{eq:VC_power_law}\,.
\end{equation}

Thus, the asymptotic $N \rightarrow \infty$ critical value $c_{0,\infty}$ can be obtained by standard least-square fitting to Eq.~(\ref{eq:VC_power_law}).  

\section{Results}
\label{sec:results}

\begin{figure}[htp]
 \includegraphics[width=0.48\textwidth]{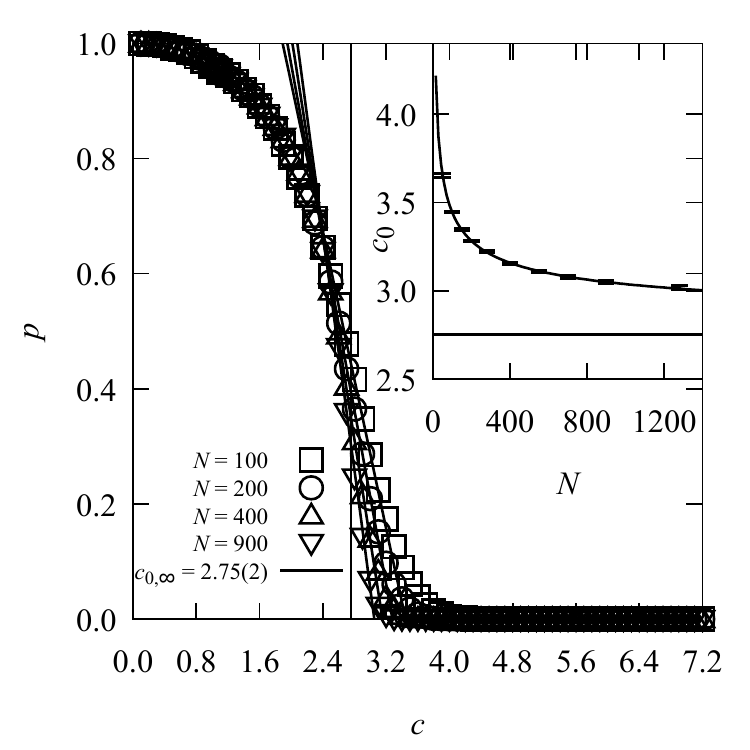}
 \caption{The fraction $p$ of complete solutions of the VC problem on ER graphs of $N$ vertices as function of the connectivity $c$, obtained with the SX algorithm. A tangent is fitted to the the steepest point of each curve, shown as straight lines. The intercept $c_0(N)$ af the tangent with the $c$-axis quantity is plotted as function of $N$, as shown in the inset. A power-law fit according to Eq.~\eqref{eq:VC_power_law}, shown as a line in the inset, results in $a = 3.8(3)$, $b = 0.35(3)$ and an asymptotic critical value $c_{\rm SX}\equiv c_{0,\infty} = 2.67(5)$, which is indicated by the vertical solid line in the main plot.  
}
 \label{fig:ER_SX_Verlauf}
\end{figure}

We performed our numerical simulation \cite{practical_guide2015} of vertex covers for the ensemble of Erd\H{o}s-Renyi graphs \cite{erdoes1960}. We analyzed graphs with the number of vertices ranging between $N=20$ and $N=3000$, respectively. For this purpose we generated graphs exhibiting exhibiting many values of the connectivity $c= 2\frac{M}{N}$ between $c = 0.1$ and at most $c=8$. For each value of $N$ and $c$ we generated an number of realizations of random graphs ranging between 20000 for the smallest sizes $N\le 500$ to at least 2000 for the largest sizes.

\subsection{Completeness of solutions}

To start, we consider the behavior of SX algorithm itself, without any refinements. The result is shown in Fig.~\ref{fig:ER_SX_Verlauf} and reveals that $p$ begins to drop already for small values of $c$. While this behavior appears to be independent of graph size $N$, the individual curves start to spread and become separated around $c \approx 2.4$. After a steep decrease of $p$, the curves flatten out, but nevertheless hit $p = 0$ on all occasions, albeit earlier for larger graphs than for smaller ones.

We have analyzed the curves as described in Sec.~\ref{sec:analysis}. The finite-size dependence of the intercepts $c_0(N)$ behave in a quite regular way and can be well fitted  by the power law of Eq.~\eqref{eq:VC_power_law}, resulting in an asymptotic value $c_{\text{SX}}\equiv c_{0,\infty} = 2.67(5)$. This is within error bars compatible with the known critical value of $c_{\text{c}} = e = 2.71828...$, where replica symmetry-breaking in the analytical calculation appears \cite{cover2000,cover-long2001}, where clustering of solutions can be observed numerically \cite{vccluster2004}, and where the percolation of the leaf-removal core occurs \cite{bauer2001}. Interestingly, at this point also the SX algorithm in combination with the cycle cutting planes stop to work successfully \cite{vc_lp2012}. Thus, the comparison with the present result seems to show that the application of cycle cutting planes leads to the same phase transition point. Nevertheless, a direct comparison of the rate $p$ of complete solutions from Fig.~\ref{fig:ER_SX_Verlauf} and Fig. 1 of Ref.~\cite{vc_lp2012} reveals, that $p$ is always higher when including the cycle cuts, as one can expect it. These figures suggest the following interpretation: For the SX algorithm alone, the point $c=e$ denotes the connectivity beyond which basically no complete solutions can be found at all, while for $c<e$, a finite fraction a graphs can be solved, but $p\to 1$ only for $c\to 0$. One the other hand, SX plus the cycle cuts allows one to find complete solutions for almost all graphs for $c<e$, in particular $p\to 1$ for about $c\le 2$.

\begin{figure}[htp]
 \includegraphics[width=0.48\textwidth]{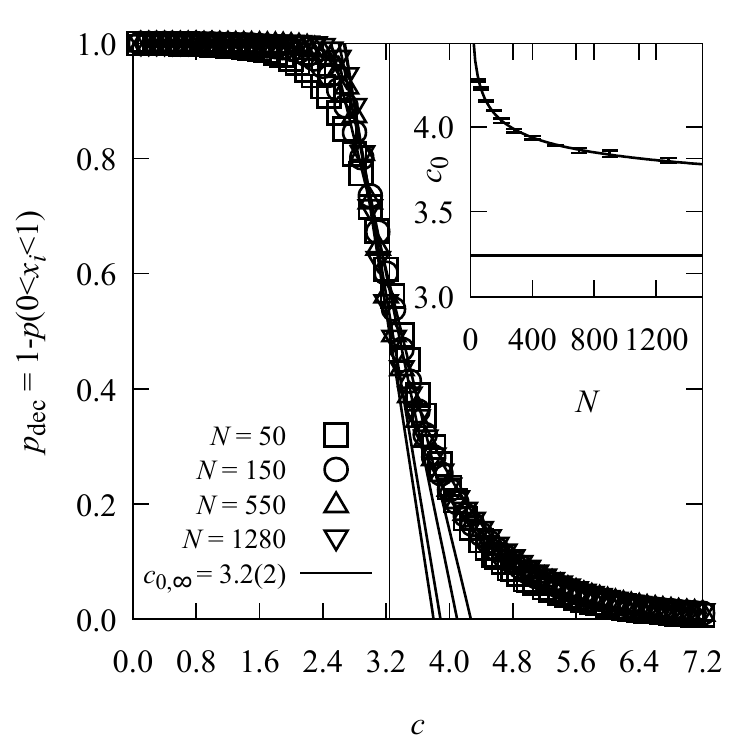}
 \caption{Fraction of decided vertices $p_{\text{dec}}$ in solutions obtained with the SX algorithm as function of connectivity $c$. The inset shows the intercepts $c_0(N)$, the extrapolation yields a critical value $c_{\text{SX}}^{\text{dec}}\equiv c_{0,\infty} = 3.2(2)$. The vertical line indicates the extrapolated critical value $c_{0,\infty}$, while the other lines show the tangents used to determine the intercepts $c_0(N)$.}
 \label{fig:ER_SX_Verlauf_entschiedene_Knoten}
\end{figure}

One might argue that the overall fraction of complete solutions is a too rigorous quantity, as a VC solution is incomplete already with only a small fraction of vertices being actually undecided. From the algorithmic point of view even obtaining a finite fraction of decided variables $x_i$, i.e., $x_i=0$ or $x_i=1$, might help towards the overall solution, because for the vertices cover problem it has been shown \cite{khuller2002} that there are always exact solutions for which the state of these variables is the same. This means, one can safely remove the nodes of these decided variables from the graph and apply a different, possibly exhaustive, algorithm for the remaining graph, which might be considerably smaller.

Therefore, we consider next the fraction of decided vertices, i.e., $p_{\text{dec}} = |\{i\in V|x_i=0 \vee x_i=1\}|/N$. The result is shown in Fig.~\ref{fig:ER_SX_Verlauf_entschiedene_Knoten}. It turns out that the fraction of decided vertices remains high for much more complex graphs, i.e. higher values of the connectivity $c$, in comparison to the result for $p$. It appears that indeed only small fractions of vertices are responsible for the rapid decrease of $p$ in Fig.~\ref{fig:ER_SX_Verlauf}. In turn, the analysis based on estimating the corresponding intercepts $c_0(N)$ with respect to the graph size $N$ yields a much higher critical value of $c_{\text{SX}}^{\text{dec}}\equiv c_{0,\infty}= 3.2(2)$. Note that the curves of $p_{\text{dec}}$ include all complete solutions as well, which enter into this averaged quantity with $p_{\text{dec}} = 1$. The actual fraction of decided vertices in graphs with \emph{incomplete} solutions may be considerably smaller. Anyway, the result means that there is an intermediate regime, where one can decide at least a fraction of the variables and the remaining problem to be solved by a complete algorithm is considerably smaller. As usual, more insight on this might be gained by examining the actual distribution of the number of decided or undecided vertices and not alone the behavior of the average, but this is beyond the scope of the current study.

\begin{figure}[htp]
 \includegraphics[width=0.48\textwidth]{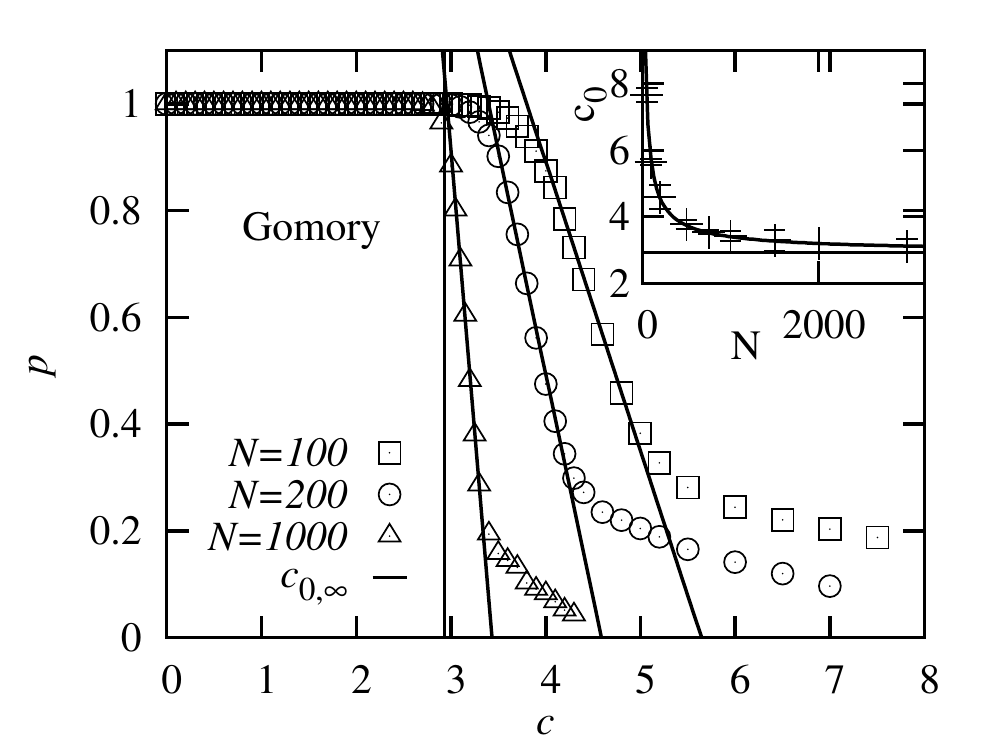}
 \caption{The fraction $p$ of complete solutions of ER graphs as function of $c$, obtained with the SX algorithm enhanced by Gomory cuts. The lines show the tangents used to determine the intercepts $c_0(N)$. The inset displays these intercepts $c_0(N)$ together with the fit according to Eq.~\eqref{eq:VC_power_law}, shown as full line, yielding a critical value of $c_{0,\infty}=2.90(2)$. The horizontal line indicates this extrapolated critical value, also shown as vertical line in the main plot.}
 \label{fig:ER_Gomory_Verlauf}
\end{figure}

Next, we consider the application of the SX algorithm with Gomory cuts. The result for $p(c)$ is shown in Fig.~\ref{fig:ER_Gomory_Verlauf}. The found curves of $p(c)$ are similar to that of the SX case in that they also approach zero. Some other properties however differ from that case: Most notably, $p(c)$ stays close to 1 over quite a range of $c$. It only starts to drop off at considerably large values around $c \approx 2.3$, also this drop is steeper than in the SX case. 

An extrapolation from the zero intercepts of the tangents on the steepest slopes yields a value of $c_\text{Gomory}\equiv 
c_{0,\infty} = 2.90(2)$, which however is well above the value found for the SX case. Thus, it seems that Gomory cuts help the SX algorithm efficiently to obtain complete solutions for connectivities beyond the aforementioned critical value at $c_{\text{c}} = e \approx 2.718$. Hence, the results offer the possibility that there exists another structural transition beyond $c_{\text{c}}=e$ which could be responsible for the transition seen with the complete Gomory-cut algorithm. However, the extent of this change of $c_\text{Gomory}$ as seen in the thermodynamic limit $N\to\infty$ may appear a bit arguable, as the extrapolated value does not lie far above that for the SX algorithm. Still, as we will see in the next section, the results for the number of necessary cuts will show that there is indeed a change of the behavior significantly above $c=e\approx 2.718$. Note that the results depend in principle on the limit $N_{\rm cut,\max}$ for the number of applied Gomory cuts. Since the Gomory cuts are complete in theory, an infinite-precision CP algorithm would always lead to a complete solution if one allows for arbitrary long running time. Anyway for some test cases in the critical region, we did not see any relevant changes when raising this limit $N_{\rm cut,\max}$ by a factor of ten. Note that if indeed raising the limit lead to any notable change, it would  rather increase of $p(c)$. This would rather increase $c_{\text{Gomory}}$ than decrease it towards $c_{\text{c}}=e$.

We have evaluated the fraction $p_{\text{dec}}$ of decided variables also for the SX + Gomory cuts approach. The figures looks similar to the previous shown results and is therefore omitted here. We have in the same way analyzed the interceptions $c_o(N)$ and fitted it to the power law Eq.~\eqref{eq:VC_power_law}. We have obtained a extrapolated critical
value $c_{\text{Gomory}}^{\text{dec}}\equiv c_{0,\infty}  = 3.48(5)$ for this case. Similar to the pure SX case, the critical point $c_{\text{Gomory}}^{\text{dec}}$ below which almost all variables can be decided is well above the critical point $c_{\text{Gomory}}=2.90(2)$ below which the problem can be completely solved. Thus, there is again an intermediate regime, where the Gomory-cut approach is able to determine many variables, leaving only a small fraction of the graph still to be solved by an algorithm which is complete in practice.

The value $c_{\text{Gomory}}^{\text{dec}} = 3.48(5)$ is near the critical value $c_{3-core}\approx3.35$ where the percolation of the 3-core appears for ER random graphs \cite{Hartmann2005}. The $q$-core of a graph is the subgraph which remains when one iteratively removes all nodes and the adjacent edges for nodes with a degree smaller than $q$. Note that for $q=3$ the transition is first order, often called ``explosive percolation''. Given this closeness, we have tried to find correlations between the set of non-integer nodes and between the 3-core by investigating realizations for $c=3.35$ and $N=1000$. But our results indicate only a small relationship: The size of the 3-core does not at all correlate with the number of non-integer variables (not shown), while it is known that the size of the leaf-removal core does \cite{vc_lp2012}, which is also confirmed by the data obtained in this study. Still, whenever there is a non-zero 3-core, the fraction of non-integer variables which are located on the 3-core is about 47\%, while the 3-core size is only about 17\% of the graph. Thus, one can say that the 3-core does not determine the number of non-integer variables, but non-integer variables are located preferentially on the 3-core.

\begin{figure}[htp]
 \includegraphics[width=0.48\textwidth]{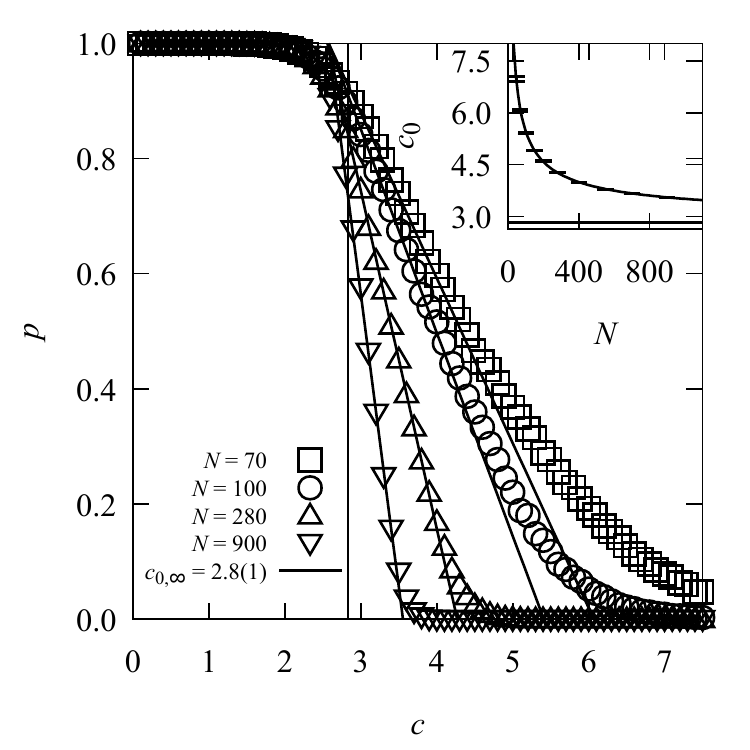}
 \caption{The fraction of complete solutions $p$ for CPLEX' MILP solver allowed to apply $\{0,\frac{1}{2}\}$ cuts. The lines show the tangents used to determine the intercepts $c_0(N)$. The inset displays these intercepts $c_0(N)$ together with the fit according to Eq.~\eqref{eq:VC_power_law}, shown as a full line, yielding a critical value of $c_{0,\infty}=2.8(1)$. The horizontal line indicates this extrapolated critical value, also shown as vertical line in the main plot.
}
 \label{fig:ER_ZeroHalf_Verlauf}
\end{figure}

Moving over to the SX algorithm with $\{0,\frac{1}{2}\}$ cuts, the result is shown in Fig.~\ref{fig:ER_ZeroHalf_Verlauf}. Note that here we rely completely on the CPLEX implementation of the cuts. Opposed to the pure SX algorithm, CPLEX then does not yield incomplete solutions here, but simply returns no solution at all and an error informing us that ``no solution exists''. Therefore, the fraction of undecided variables can not be evaluated. Anyway, the result of $p(c)$
displays similar qualities compared to Gomory cuts, including the tendency to remain near $p(c) \approx 1$ until $c \approx 2.3$. The critical point for the disappearance of complete solutions is determined by the fit to Eq.~\eqref{eq:VC_power_law} and yields the extrapolated value $c_{0,\frac{1}{2}} \equiv c_{0,\infty} = 2.8(1)$. This is statistically compatible with the critical values obtained for the SX approach alone, the SX + cylce cuts algorithm \cite{vc_lp2012} and the critical value $c_{\text{c}}=e\approx 2.718$. Since, as discussed, the previously used cycle cuts are a, apparently powerful, subset of the zero-half cuts, it appears reasonable that the critical point is comparable, if not equal.

It might be tempting to allow Gomory and $\{0,\frac{1}{2}\}$ cuts at the same time. We actually carried out such calculations, yet it turned out that the found value for $c_{0,\infty}$ did not go beyond those found for these two CP methods alone, respectively. Therefore, we omit these results here, and conclude that the applicability of cutting planes is controlled by the set of ``most-powerful'' cuts and the corresponding critical points determined by the related structural changes of the graph ensemble.

\subsection{Computational hardness}

The next question we want to address is whether there is a connection between the behavior of the solvability and the typical computational hardness. Regarding the application of the pure SX algorithm, this question is not so interesting, because it is known that the SX algorithm typically runs always in polynomial time \cite{papadimitriou1998}, although there is not formal polynomial bound for the worst-case running time. Note that the so called ellipsoid algorithm solves LP in guaranteed worst-case polynomial time, but it is more complicated and slower than SX.

\begin{figure}[htp]
 \includegraphics[width=0.42\textwidth]{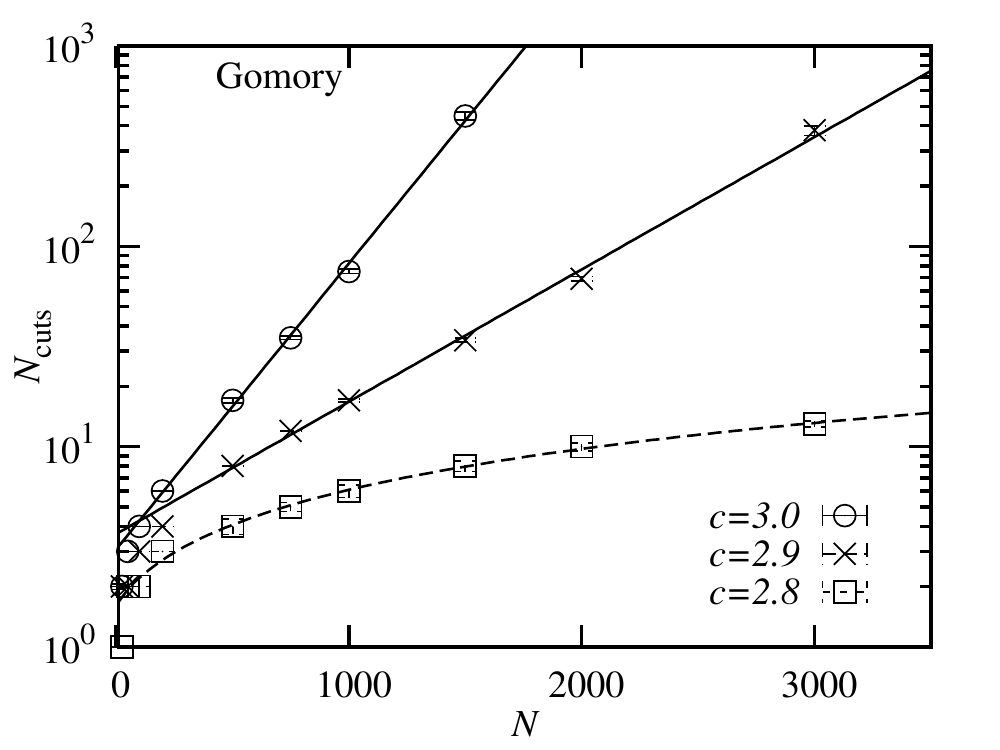}
 \caption{The 70\% percentile of the number $N_{\rm cuts}$ of Gomory cuts until a complete solution is found as a function of the system size $N$, for some values of the connectivity $c$. For $c=2.8$, the data can be well fitted by a power law with exponent $b=0.86(7)$ (broken line), while for $c=2.9$ and $c=3.0$ the behavior is compatible with an exponential (full lines).}
 \label{fig:time:gomory:N}
\end{figure}

Thus, we consider next the case of the Gomory cuts. We investigated the \emph{typical} number $N_{\rm cuts}$ of cuts, as usual when considering running times, since it will not be influenced by statistical outliers which are very hard to solve.
In particular, we evaluated the $r=0.7$ percentile. Thus, 70\% of all problem instances require a number of cuts smaller or equal to the  shown value. Note the we could have used the median, i.e., the 0.5 percentile instead, but here the finite-size dependence is not very strong since about half of the problem instances require only few cuts. This means, for the
median it is much harder to distinguish between a polynomial and an exponential growth. Note that when taking the percentile also instances which can not be solved within the selected maximum number of added cuts have no influence, if the fraction of solved instances is larger than $r$. This is also the reason why we have not chosen a larger percentile $r$, because then the number of unsolved instances would make more points invalid: For $c=2.9$ and $N=3000$, the fraction of unsolved instances has already grown to $0.25$.

For small values of $c$, the problem is typically solved with no or very few additional cuts. Thus, we concentrate on the case of $c$ near the critical value $c_{\text{c}}=e$. In Fig.~\ref{fig:time:gomory:N} the typical number $N_{\rm cuts}$ of
Gomory cuts as a function of $N$ is shown for three significant values of $c$. Please note the logarithmic scale for $N_{\rm cuts}$. For $c=2.8$, the data exhibits a curvature and a fit to a power law $N_{\rm cut}(N)=N_{\rm cut,0}+\alpha N^\beta$ works very well, with $N_{\rm cut,0}=1.6(2)$, $\alpha=0.012(7)$, and a small $\beta=0.86(7)$. This shows that the problem is typically polynomial here, even slightly beyond $c=e$. Note that, as additional test, for $c=2.8$ also the \emph{average} number of Gomory cuts (not shown) exhibits a polynomial behavior with power as a function of the system size $N^\beta$ with $\beta=1.29(4)$.

On the other hand for $c=2.9$ and $c=3.0$, the data follows a straight line and fits well to an exponential $N_{\rm cut}(N)=\gamma\exp(\delta N)$, with $\gamma=3.7(2)$, $\delta=0.015(1)$ for $c=2.9$ and $\gamma=3.1(4)$, $\delta=0.033(1)$ for $c=3.0$. This renders the problem typically exponentially hard for $c\ge 2.9$. These results are compatible with the above determined critical threshold $c_\text{Gomory}=2.90(2)$ below which typically complete solutions are found by this approach. Note in particular the fact that the number of cuts at $c=2.8$ increases clearly polynomially, at least for the system sizes we can access, confirms that that the critical value $c_{\text{Gomory}}$ for the Gomory cuts is separated from the well known value $c_{\text{c}}=e$ where RSB appears.
 
\begin{figure}[htp]
\includegraphics[width=0.48\textwidth]{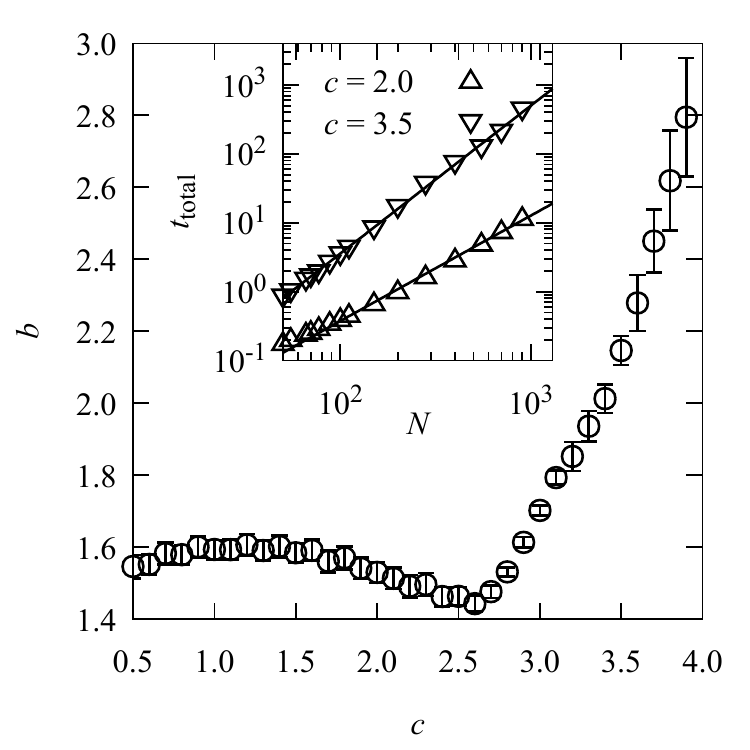}
 \caption{Calculation times 
$t_{\text{total}}$ needed to obtain the full ILP solution of CPLEX. The inset shows that the time exhibit a power-law dependence on the graph size $N$ according to $t_{\text{total}} = a\cdot N^b$. The main plot shows the resulting exponent $b$ as function of the average connectivity $c$. $b$ starts to increase drastically once the connectivity $c$ exceeds the critical value at $c_{\text{c}} = e$.
\label{fig:ER_MIP_Rechenzeit}
}
\end{figure}

The CPLEX package which we have used offers also a complete solver for ILPs through a number of means beyond those considered here in detail. Among these are branching, all kinds of different CP methods and not further specified heuristics, all of which are even enabled per default. This however comes at the price of a massive numerical effort, which can be estimated directly from the 
solver's ``tics''. This is stated by CPLEX and measured versus a given counter
during the execution of the code and can therefore be interpreted as deterministic and somehow independent of the respective computer system. Thus, we use this number of tics as calculation time $t_{\text{total}}$ of the complete solver and study its dependence on the graph size $N$ at given values of the graph connectivity $c$. It turns out that the behavior of $t_{\text{total}}$ is well described through a power-law dependence, i.e. $t_{\text{total}} \propto N^b$ in
any case as exemplified in the inset of Fig.~\ref{fig:ER_MIP_Rechenzeit}. Note that we observe polynomial behavior for all values of $c$, even well beyond $c_{\text{c}}=e$. This shows that the algorithm is in the range of accessible system sizes empirically very powerful. Still, it can not be excluded that for much larger sizes an exponential behavior becomes visible for large enough connectivity $c$. Still, for low values of $c$, the exponents $b$ revolve around $b \approx 1.5$, in the range of $b = 1.4...1.6$. However, the main plot of Fig.~\ref{fig:ER_MIP_Rechenzeit} exposes a minimum at $c \approx 2.6$.
The reason for the slight decrease beyond $c\ge 1$, instead of a slight growth, is probably that CPLEX starts to use other means than the cutting planes studied here, e.g., activates branching, since this is actually more efficient. Beyond $c\approx 2.6$ an excessive growth of $b$ can be observed. Thus, although the typical behavior is polynomial for the complete CPLEX solver in the studied regime, the combination of algorithms CPLEX uses to attack VC is again mostly influenced by the structural change of the ER random graphs near $c_{\text{c}}$, highlighting the importance of this critical point.

Note that we could have measured also the running time for the zero-half cuts through the number of CPLEX tics. Nevertheless, the behavior of the zero-half cuts is very similar to the cycle cuts, as explained. Since the empirical computational hardness of the cycle cuts has already been studied \cite{vc_lp2012} and did also show a pronounced change at $c_{\text{c}}=e$, we do not expect different results here and thus did not consider the empirical time complexity of the zero-half cuts.

\subsection{Whiteness of vertex-cover solutions \label{sec:whitening}}

The intention behind investigation of whitening in VC solutions is the idea that the aforementioned transition at $c_{\text{c}} = e$ might also affect quantities related to white or frozen vertices of these solution. The whitening procedure can only be applied properly on correct, complete solutions of the VC problem on a given graph. Therefore, for the corresponding simulations we simply utilized all options CPLEX has to offer, heedless of the drastic increase of computation time for larger connectivities. Still, it was sufficient to consider only graphs up to $c = 4$ where we were able to study graphs with $N\le 700$.

Our results show that, even with definition of isolated vertices as white, full-white solutions are rare and only occur for graphs of small size $N$ and low connectivity $c$, so the VC problem is generally characterized by largely frozen variables and only a certain fraction of white vertices or clusters of these. 

\begin{figure}
 \includegraphics[width=0.48\textwidth]{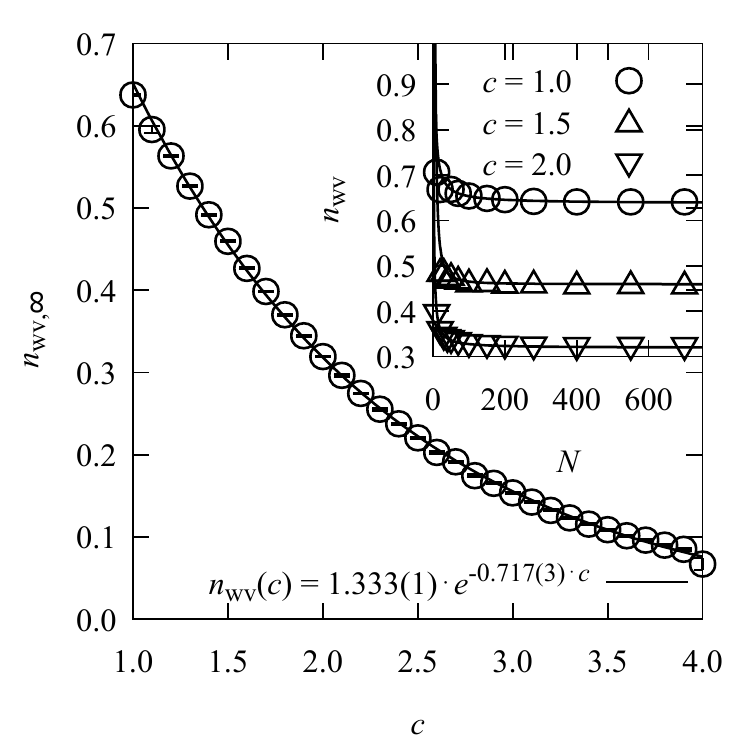}
 \caption{The extrapolation $n_{\text{wv},\infty}$ of the relative number of white vertices adheres smoothly to an exponential decay over graph connectivity $c$. As shown in the inset, these extrapolated values result from an asymptotic decay of this quantity over graph size $N$ at fixed values of $c$.}
 \label{fig:ER_Whitening_relative_Anzahl_weisser_Knoten}
\end{figure}

First, we considered the size dependence of the number $N_{\text{wv}}(N)$ of white vertices. For a given connectivity $c$, the relative quantity $n_{\text{wv}}=N_{\text{wv}}/N$ converges asymptotically to a constant value according to a power-law like Eq.~\eqref{eq:VC_power_law}, i.e., $n_{\text{wv}} = n_{\text{wv},\infty}+a N^{-b}$, see inset of Fig.~\ref{fig:ER_Whitening_relative_Anzahl_weisser_Knoten}. These extrapolated values $n_{\text{wv},\infty}$ as function of $c$ are shown see main plot in the figure. No particular influence of the transition at or near $c_{\text{c}} = e$ or elsewhere is visible. The asymptotic fraction $n_{\text{wv}}$ of white vertices matches well an exponential decrease as function of $c$. This resembles  the exponential decrease $e^{-c}$ of the isolated vertices, but not as fast. Already at $c=1$, the fraction of isolated vertices is about 0.37, well below $n_{\text{wv},\infty}\approx 0.64$.

\begin{figure}
 \includegraphics[width=0.48\textwidth]{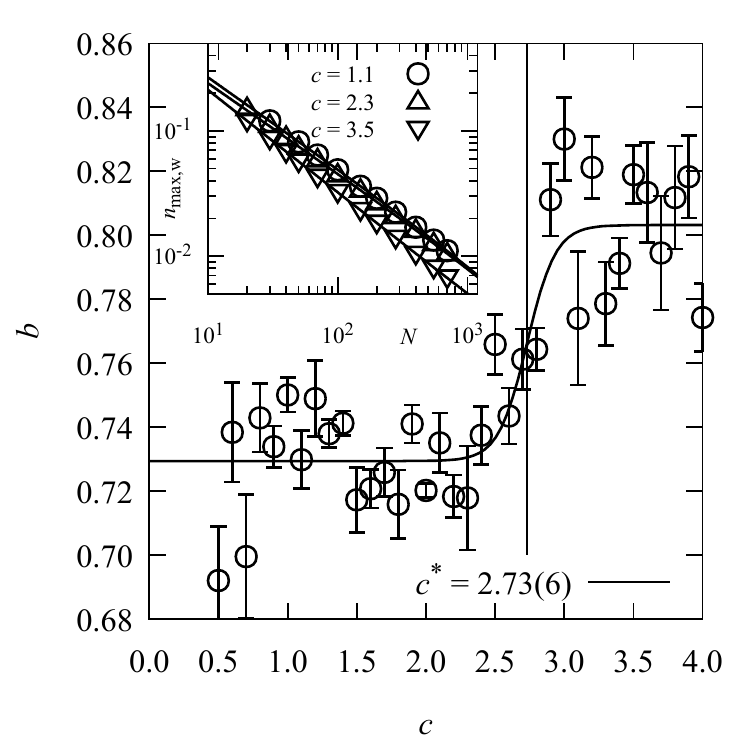}
 \caption{Inset: relative size $n_{\text{max,w}}$ of the largest white cluster, indicating a power law behavior $N^{-b}$. Main plot: dependence of the exponent $b$ on the connectivity $c$ which demonstrates the criticality of the VC by ``jumping'' from a lower value of about $b \approx 0.73$ to a higher value of $b \approx 0.8$ once the connectivity $c$ rises. The line displays the sigmoid function $\text{sig}(c)$ fitted to the data, with a
center value  $c^* = 2.73(6)$ (marked by the vertical line).}
 \label{fig:ER_Whitening_relative_Groesse_groesster_weisser_Cluster}
\end{figure}

Still,  by taking a closer look, the RSB transition at $c_{\text{c}}=e$ is visible in the data. In the inset of Fig.~\ref{fig:ER_Whitening_relative_Groesse_groesster_weisser_Cluster} the relative size $n_{\text{max,w}}=N_{\text{max,w}}/N$ of the largest white cluster is shown as a function of the number $N$ of nodes. The data follows well a power law with exponent $b$ for all values of $c$. In the main plot of  Fig.~\ref{fig:ER_Whitening_relative_Groesse_groesster_weisser_Cluster} the results for $b$ from fits to power laws are shown as function of the connectivity $c$. For low values of $c$, $b$ revolves around low values of ca. $b \approx 0.73$. With $c$ increasing above the threshold around $c_{\text{c}} = e$, the value of the exponent then moves to a regime of considerably larger values around $b \approx 0.8$. It is also possible to fit a stretched and shifted sigmoid function $\text{sig}(c) = b_0+\frac{a}{1+\exp (c-c^*)}$ to the course of $b(c)$ to model this transition and yield a center point of $c^* = 2.73(6)$, which which is compatible with the RSB transition at $c_{\text{c}} = e$ within the error range.

For some other quantities obtained through fits similar changes near $c_{\text{c}}=e$ are visible. For example, when considering the average size of all clusters except the largest one, and fitting power laws to the $N$-dependence, the prefactor of the resulting power law shows a decrease by about 15\% (not shown) near $c=c_{\text{c}}$, but the signature is a bit weaker as compared to the preceding case. Next, when analyzing the number of white clusters as function of system size, the approach to the limiting value is from below for about $c\le 2.6$ while it is from above for larger values of $c$. Nevertheless, such pronounced changes as function of $c$ are not visible for all parameters involved in the finite-size behavior of parameters
related to whitening. For example, we also considered the exponent $b$ obtained in fitting $n_{\text{wv}}(N)$, as shown in the inset of Fig.~\ref{fig:ER_Whitening_relative_Anzahl_weisser_Knoten}. Here the value of  $b$ is up to a factor of 2 larger for smaller values of $c\le 2$ as compared to large values $c\ge 3$ (not shown). But the data for $b$ is rather noisy and the decrease seems to happen rather near $c=2$ than near $c_{\text{c}}=e$.


\section{Conclusions \label{sec:conclusions}}

We have considered the NP-hard combinatorial optimization problem vertex cover on Erd\H{o}s-Renyi random graphs with connectivity $c$. We applied a
represention as a linear program  and sought to solve this through various variants of the simplex algorithm in particular enhanced with Gomory cuts. The main motivation behind this was the question how the structure of the ER graphs influences the ability of the SX algorithm augmented with a theoretically complete set of cutting planes to obtain integer solutions at the expense of computational effort. Also, due to numerical limitations originating from finite number precision, well known instabilities of Gomory cuts may lead to regions in graph space, where in practice no solutions can be obtained at all.

By applying the CPLEX library for LPs, we could study rather large graph sizes of up to $N=3000$ nodes. We have found transitions between solvable and unsolvable phases for SX alone and for SX+Gomory cuts. By analyzing the finite-size dependence of the interceptions $c_0(N)$ of tangents of the solution probability $p(c)$, we determined critical points by extrapolation $N\to\infty$. For the pure SX algorithm, the obtained value $c_{\rm SX}=2.67(5)$ is well compatible with the critical point $c_{\text{c}}=e\approx 2.718$ where the onset of RSB is located, a hierarchical clustering of the solution sets 
can be observed, and the ER ensemble exhibits a percolation transition of the leaf-removal core. Interestingly, the same analysis for the SX+Gomory cuts approach yields a significantly higher value $c_{\text{Gomory}}=2.90(2)$. The analysis of the number $N_{\text{cuts}}$ of necessary Gomory cuts supports the finding that this easy-hard transition is distinct from the RSB transition: For $c\le 2.8$, $N_{\text{cuts}}$ grows like a power law with the system size, while for $c\ge 2.9$ a clear exponential growth is visible. Thus, it seems that the structure of ER random graphs exhibits another, to our knowledge yet unknown, change of structure for $c\in[2.8,2.9]$. Nevertheless, since we study only graphs of finite sizes numerically, we cannot exclude that the apparent polynomial growth of the running time we have observed for $c=2.8$ turns into an exponential growth at much larger system sizes, but this appears unlikely to us, given the clean fits we have observed. Anyway, it would be very interesting if our results motivate further analysis which confirm that VC indeed exhibits more than one transition with respect to the computational complexity, similar, e.g., to the rich behavior of the SAT problem.

Concerning the technical implementation of Gomory cuts, it should be noted that they require very high numerical precision, since they rely on an accurate computational representation of rational numbers. CPLEX actual operates with a finite numerical percision. At least, cross-checking with other publicly available LP solvers such as \texttt{lp\_solve},\cite{LPSOLVE} \texttt{PuLP}\cite{PULP} or \texttt{Python MIP}\cite{PYTHONMIP} might give further insight to this issue, to see whether the observed easy-hard transition at a point above $c_{\text{c}}=e$ is present for other solvers as well.

We have also studied the fraction $p_{\text{dec}}$ of decided variables which results form the calculation of partial solutions. This is interesting, because for VC all integer-valued variables within a incomplete solution can be left out before proceding with a complete algorithm. We have found also
transitions from a low-connectivity phase where $p_{\text{dec}}>0$, but possibly $p_{\text{dec}}<1$ to a high-connectivity phase where $p_{\text{dec}}=0$. Interestingly, by again analyzing the finite-size behavior of the intercepts of tangents, we obtained for the SX and for the SX+Gomory cases of transition points 
around $c=3.3$ which are located even more above the known RSB transition point $c_{\text{c}}=e$. Our analysis revealed a weak relation to the appearance of the 3-core, but whether undecided variables exist at all in the solution seems basically to be independent of the existence of a 3-core.

Our results show that the dynamics of LP based algorithms are harder to understand. The reason is probably that they operate outside the space of feasible solutions, in contrast to branch-and-bound algorithms, message-passing techniques  or approaches based on random walks, where only feasible solutions are considered. Thus, LP-based algorithms deserve a closer look by physicists in the future, in particular because such algorithms are applied in practical industrial applications.

Furthermore, it turned out that $\left\{0,\frac{1}{2}\right\}$ cuts cannot be seen an alternative to Gomory cuts. The obtained phase transition rather coincides with that observed for the pure SX approach and the one seen in previous work for SX + cycle cuts. This seems plausible, since we argued that they technically are largely identical to the model-specific CP approach of \cite{vc_lp2012}, which considered particular constraints for vertex cycles of odd length within the ER graph.

Regarding whitening of the VC solutions, the findings for this problem differed well from those reported for SAT models in that the VC problem hardly ever showed full whitening. This however comes as no big surprise, as we had to modify the whitening procedure specifically for this problem since the original whitening procedure for SAT tended to mark the whole interconnected cluster of the graph as white. The asymptotic $N\to\infty$ fraction of white vertices showed no pronounced change as a function of the connectivity $c$, just a smooth exponential decay. But instead, we found out that the criticality exposes itself in other properties related to whitening, namely in the exponent which controls the finite-size behavior of the size of the largest white cluster.

To summarize, whitening for VC, at least in the version we tried, displays visible effects related to the clustering phase transition at $c_{\text{c}}=e$ only within a deeper scaling analysis of the data. A more direct access to the clustering structure, e.g., by looking at neighboring solutions in solution space \cite{vccluster2004}, seems so far more capable to provide insight into the solution landscape for VC.

Concerning further work, it should be stressed that Gomory cuts are in principle able to solve any combinatorial optimization problem, by considering relaxed LPs, i.e., configurations which start non-feasible but optimal and are moved iteratively into the space of feasible solution. This should motivate more research, seeking a better understanding of the relation between computational complexity, problem instance properties and solution-space structure. Here, checking the VC problem through other LP solvers with Gomory cuts might add clarity, in particular for solvers which work with true infinite-precision rational number representations. Also it would certainly be worthwhile to consider other problems, like coloring, SAT, number partitioning or the traveling salesperson problem, to see whether known easy-hard transitions are visible when studying the behavior of Gomory cuts.

Also inspection of typical properties of complete solutions of VC and other optimization problems might give insight to similar CP methods. It could be even worthwhile to utilize machine-learning based methods for the selection of cutting pleanes, since such approaches to the general VC problem have already been considered in other ways \cite{Ramanujam1988,Evans1998,Chen2004}. 

Finally, the work on the relation between out-of-configuration space LP cutting plane approaches and the structure of combinatorial optimization problems has been numerical so far, to our knowledge. It would therefore be very desirable if analytical approaches were attempted to better understand how the structure of the solution space affects such algorithms.

\begin{acknowledgments}
We acknowledge personal support by Christoph Norrenbrock. The simulations were performed at the the HPC cluster CARL, located at the University of Oldenburg (Germany) and funded by the DFG through its Major Research Instrumentation Program (INST 184/157-1 FUGG) and the Ministry of Science and Culture (MWK) of the Lower Saxony State. 
\end{acknowledgments}


\bibliography{VC_Cuts.bib}

\end{document}